\documentclass[11pt]{article}
\usepackage{moriond,epsfig}

\def\Journal#1#2#3#4{{#1} {\bf #2}, #3 (#4)}

\def\be{\begin{equation}}
\def\ee{\end{equation}}
\def\bea{\begin{eqnarray}}
\def\eea{\end{eqnarray}}

\def\OmM{\Omega_{\rm M}}
\def\deltac{\delta_{\rm c}}
\def\Dg{D_{\rm g}}
\def\fx{f_{\rm x}}
\begin{document}
\vspace*{4cm}
\title{THE XMM--NEWTON $\Omega$ PROJECT}

\author{ J.G.~BARTLETT$^{1,2}$, N.~AGHANIM$^3$, M.~ARNAUD$^4$,
J.--PH.~BERNARD$^3$, A.~BLANCHARD$^1$, M.~BOER$^5$, D.J.~BURKE$^6$, 
C.A.~COLLINS$^7$, M.~GIARD$^5$, D.H.~LUMB$^8$, S.~MAJEROWICZ$^4$, 
PH.~MARTY$^3$, D.~NEUMANN$^4$, J.~NEVALAINEN$^8$, R.C.~NICHOL$^{9}$,
C.~PICHON$^{10}$, A.K.~ROMER$^{9}$, R.~SADAT$^1$, C.~ADAMI (associate)}

\address{1.~Observatoire Midi--Pyr\'en\'ees, Toulouse, France\\
2.~CDS, Strasbourg, France\\
3.~Institut d'Astrophysique Spatiale, Orsay, France\\
4.~SAp, CEA, Saclay, France\\
5.~CESR, Toulouse, France\\
6.~CfA, Cambridge, USA\\
7.~Liverpool John Moores University, UK\\
8.~Astrophysics Division, ESA/ESTEC\\
9.~Carnegie Mellon University, Pittsburgh, USA\\
10.~Observatoire de Strasbourg, Strasbourg, France\\
}

\maketitle\abstracts{The abundance of high--redshift galaxy clusters depends 
sensitively on the matter density $\OmM$ and, to a lesser extent, 
on the cosmological constant $\Lambda$.  Measurements of this 
abundance therefore constrain these fundamental cosmological parameters,
and in a manner independent and complementary to other methods, such
as observations of the cosmic microwave background and distance measurements.
Cluster abundance is best measured by the X--ray temperature
function, as opposed to luminosity, because temperature and mass
are tightly correlated, as demonstrated by numerical simulations.
Taking advantage of the sensitivity of XMM--Newton, our Guaranteed Time 
program aims at measuring the temperature of the highest redshift 
($z>0.4$) SHARC clusters, with the ultimate goal of constraining
both $\OmM$ and $\Lambda$.
}

\section{Cluster abundance and Cosmology}

        In standard models, structures form from the collapse of density
perturbations described by a Gaussian random field (in the linear regime).
An object collapses once the density contrast $\delta\equiv (\rho-\bar{\rho})/
\bar{\rho}$ (where $\rho$ is the density field) reaches a critical value
$\sim 1$.  The abundance of such regions will reflect the Gaussian nature
of the perturbations, as will the mass function giving the
number density of objects as a function of mass $M$ and redshift $z$.  Using
simple statistical arguments of this kind, Press \& Schechter~\cite{ps} (PS)
suggested the following formula for the mass function
\begin{eqnarray}
\frac{dn}{d\ln M} & = & \sqrt{\frac{2}{\pi}}\frac{\bar{\rho}}{M}
     \nu(M,z) \left|\frac{d\ln \sigma}{d\ln M}  \right| 
     e^{-\nu^2/2} \\ 
\nonumber
\\
\nonumber
\nu(M,z) & \equiv & \deltac/\sigma(M,z)
\end{eqnarray}
where $\deltac$ is a (weakly) cosmology--dependent threshold ($\sim 1.68$)
and $\sigma(M,z)$ is the density perturbation amplitude at scale $M$.
We see the underlying statistical nature of the perturbations in
the Gaussian cut--off at the high--mass end.
Expressions for the mass function found in large numerical simulations 
differ somewhat from the PS form, but are well described
by similarly simple analytic expressions~\cite{jenkins}.  Clusters reside
on the high--mass tail (where $\sigma<1$) and their abundance
at any $z$ is therefore highly sensitive to 
\begin{equation}
\sigma(M,z) = \sigma(M,z=0) \Dg(z;\OmM,\Lambda)
\end{equation}
Here $\Dg$ is the growth factor for linear perturbations, which
depends on the cosmology ($\OmM,\Lambda$).  
Perturbations {\em freeze--out}, i.e.,
slow their growth rate, in low--density models when the expansion
becomes dominated by either the curvature term or a cosmological
constant; $\Dg$ thus depends primarily on $\OmM$, and to a 
lesser extent on $\Lambda$.  The presence of a cosmology--dependent
factor in the exponential of the mass function implies that
cluster abundance is an effective way to constrain these
cosmological parameters.  Constraints obtained in this 
manner are complementary to, for example, those
found by observations of supernovae type Ia or by
measurements of cosmic microwave
background anisotropies that essentially rely
on a determination of cosmological distance (luminosity
or angular--size distances). 

\section{X--ray Temperature}

          Strictly speaking, one requires the abundance of clusters
as a function of their mass, a difficult quantity to measure 
directly.  In practice, one seeks a direct observable 
that is closely related to virial mass.  Lensing surveys
would seem the most suited to the task, as the effects
of lensing are of course directly related to mass 
(although projected along the line--of--sight).
Among X--ray observables, temperature is a much
more robust quantity than the luminosity, the latter
depending on the density profile of the intracluster
gas whose physics is currently difficult
to model.  The X--ray temperature, on the other hand,
is expected to be tightly correlated with virial mass,
an expectation borne out by numerical simulations~\cite{evrard,bn}.

\begin{figure}
\rule{5cm}{0.2mm}\hfill\rule{5cm}{0.2mm}
\hspace*{2cm}
\psfig{figure=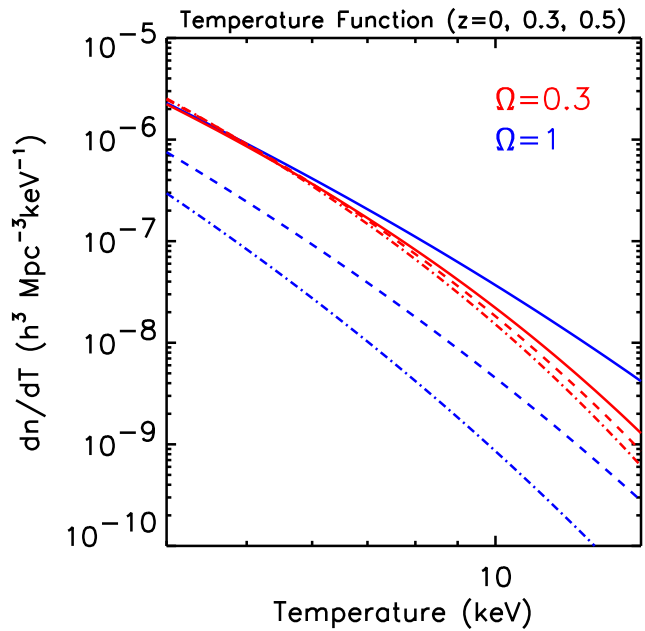, height=9cm}
\caption{The predicted X--ray temperature function at redshifts
$z=0$ (solid), 0.3 (dashed) and 0.5 (dot--dashed) for a critical
(blue) and an open (red) model ($\Lambda=0$), 
both fitted to the local ($z=0$) temperature
abundance.
\label{fig:dndT}}
\vskip 1.0cm
\rule{5cm}{0.2mm}\hfill\rule{5cm}{0.2mm}
\end{figure}

     Cluster abundance, and its evolution, is thus well
measured by the X--ray temperature function $dn/dT(T,z)$.
With a calibrated $T-M$ relation, the 
mass function is easily translated into a temperature
function that may be compared with observations. 
The exact $T-M$ relation to use is of course a critical
issue, one that may be addressed, for example, using 
numerical simulations, or directly from detailed observations
that determine both cluster mass and temperature, e.g.,~\cite{jukka}.
Figure~\ref{fig:dndT} compares the predictions for a critical
and an open model, both normalized to the present--day,
observed $dn/dT$.  Evolution towards higher redshift
is strikingly different in the two models, illustrating the power 
of measurements of cluster abundance at $z>0$ as a cosmological
probe\cite{ob1}.  This probe has been applied by numerous 
authors, yielding a variety of results on 
$\OmM$~\cite{ob2,h97,bf,bb,sbo,borgani,eke,markevitch,reichart,vl}.

     Use of the cluster abundance as a cosmological probe
requires a well--controled sample with temperature 
measurements.  There are several estimates of the 
local temperature function at $z=0$\cite{ha,edge,b2000}.  
For many years,
the EMSS~\cite{emss} played the central role for studies at $z>0$; 
Henry~\cite{h97} used this sample to find the temperature
function at $z\sim 0.3$, which is still
the most distant temperature function determinated to date. 
In addition, there are now several 
X--ray selected catalogs based on serendipitous
cluster detections in ROSAT pointings that 
are being used for cluster evolution
studies.  Efforts are underway to obtain temperatures
for these samples using the new X--rays satellites,
Chandra and XMM--Newton, but as yet there are no
conclusive results.  

     Temperature measurements are difficult
to obtain because they require many photons to construct
an X--ray spectrum; this is of course the reason
the temperature function is still only poorly known
at $z>0$.  Another avenue to the cluster abundance at high
redshift is to apply a luminosity--temperature relation
to a flux--limited sample, thereby obtaining either
the temperature function, or a redshift distribution
at given temperature.  The advantage is that the 
luminosity--temperature relation may be determined
over a range of redshifts with few objects, and
then applied to the much larger parent catalog~\cite{sbo}.

     In either case, whether one wishes to directly determine
the temperature function, or use a constrained luminosity--temperature
relation on a large flux--limited sample, temperature measurements
for a number of clusters at high redshift are needed.  This
is the goal of our XMM--Newton $\Omega$--project based on
the SHARC cluster sample.

\begin{center}
\begin{figure}
\rule{5cm}{0.2mm}\hfill\rule{5cm}{0.2mm}
\hspace*{4cm}
\psfig{figure=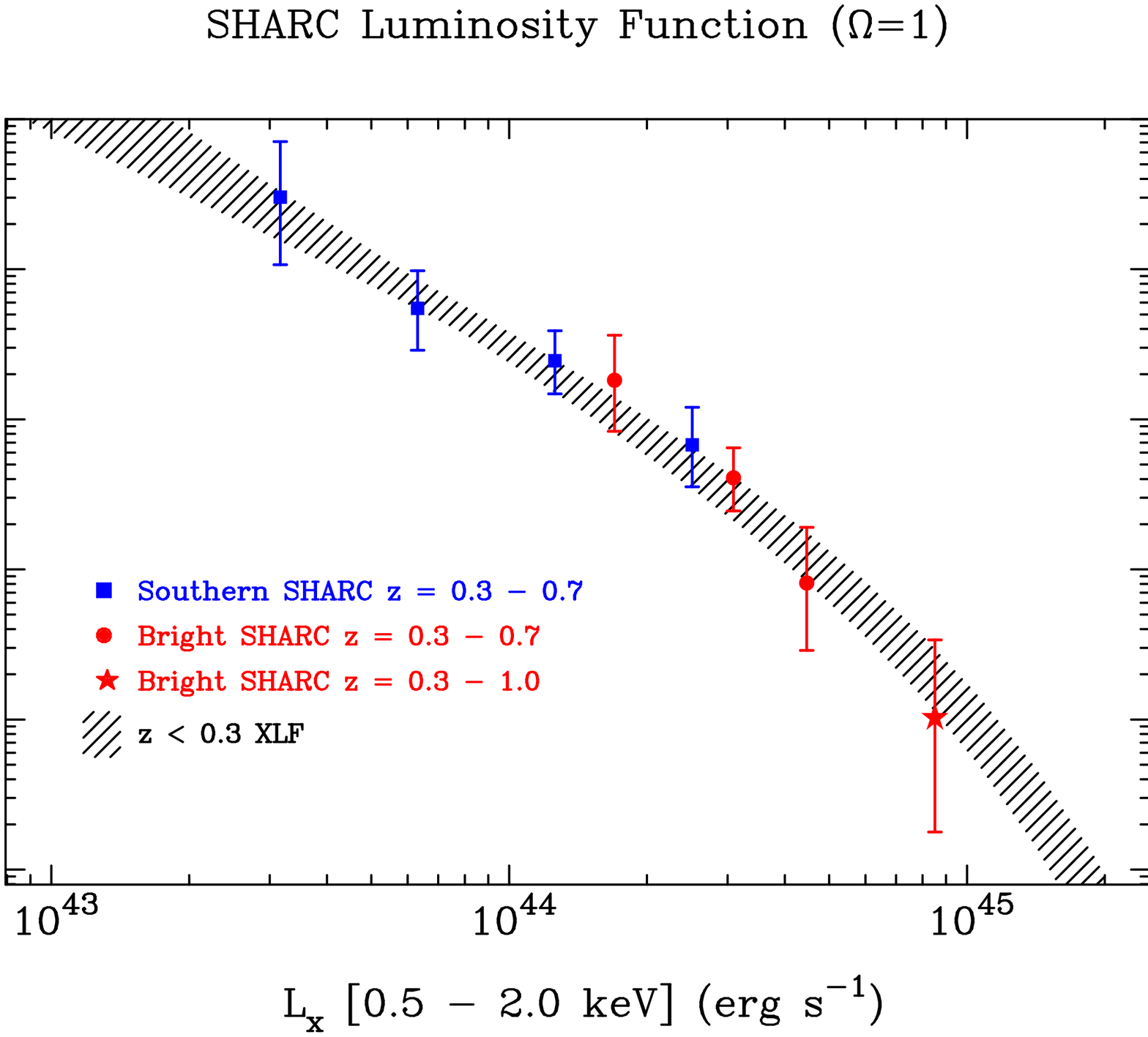,height=9cm}
\caption{The SHARC luminosity function.
\label{fig:SHARCLumFunc}}
\vskip 1.0cm
\rule{5cm}{0.2mm}\hfill\rule{5cm}{0.2mm}
\end{figure}
\end{center}

\section{The SHARC Sample}

     The Serendipitous High--redshift Archival ROSAT Cluster
(SHARC) survey consists of sources found to be extended by
a wavelet analysis in ROSAT pointings.  The catalog
consists of objects found in two separate surveys:
the Deep SHARC~\cite{deepsharc1,deepsharc2}, covering 17 square degrees
in the South to a flux limit of $\fx[0.5-2{\rm keV}] >
4\times 10^{-14}$~ergs/s/cm$^2$; and the Bright 
SHARC~\cite{brightsharc1,brightsharc2}, covering 178 square degrees to 
$\fx[0.5-2{\rm keV}] > 1.5\times 10^{-13}$~ergs/s/cm$^2$.
This two--fold strategy yields a cluster catalog that
straddles $L^*$ over $0.2<z<0.8$ (see Figure~\ref{fig:SHARCLumFunc}).
The selection function for the SHARC has been extensively
studied~\cite{adami}.

\section{The XMM--Newton Project}

     As mentioned, the difficulty in obtaining temperatures
for high--redshift clusters lies in collecting enough 
photons.  The large collecting area of XMM--Newton
makes it an ideal instrument for the task.  Our 
Guaranteed Time (GT) program aims to establish the
luminosity--temperature relation and to estimate
the temperature function at the highest possible 
redshifts.  To this end, we will observe 7 of the
most distant SHARC clusters, all at $z>0.4$ (median
of $z=0.5$), to find their temperatures to $\sim 10$\%
accuracy.  Our consortium consists of GT holders
from the EPIC, SOC and SSC XMM teams, and the
SHARC team.  The total observing time on the GT program
amounts to 260~ksecs for the 7 objects, plus an additional
cluster from another GT program obtained by mutual
agreement; the program will continue on Guest
Observing (GO) time with 160~ksecs already allocated
after the first AO.  

     With the 7-8 temperature measurements on the GT program, 
we will directly construct the temperature function at 
the highest redshifts yet reached.  As the ability to 
distinguish models improves rapidly with redshift, due
to the fact that clusters lie ever farther out on the
Gaussian tail at early times, we hope to significantly
improve constraints on $\OmM$ and $\Lambda$ from 
this key cosmological probe.  In addition, 
the results will be based on a cluster catalog
entirely independent of the workhorse EMSS, and therefore
with different possible systematics, which will help
to control possible hidden systematics in the method.

\section*{Acknowledgments}
Our thanks to the organizers for an enjoyable and exciting 
meeting, and to the local staff for their constant attention
and care.


\end{document}